\documentstyle[preprint,aps,epsbox]{revtex}

\makeatletter

\def\lsim{\compoundrel<\over\sim}
\def\compoundrel#1\over#2{\mathpalette\compoundreL{{#1}\over{#2}}}
\def\compoundreL#1#2{\compoundREL#1#2}
\def\compoundREL#1#2\over#3{\mathrel
  {\vcenter{\hbox{$\m@th\buildrel{#1#2}\over{#1#3}$}}}}
\makeatother

\newcommand{\mapright}[1]{%
	\smash{\mathop{%
	\hbox to 1cm{\rightarrowfill}}\limits^{#1}}}

\newcommand{\ssx}{
\sqrt{
\frac{m_u(m_t+m_u)(m_t-m_c-m_u)}
{(m_c+m_u)(m_t^2-m_t m_c -m_u^2)}}}

\newcommand{\ssy}{
\sqrt{
\frac{m_u(m_t-m_u)}{m_t^2-m_t m_c - m_u^2}}}

\newcommand{\sx}{
\sqrt{
\frac{m_d(m_b+m_d)(m_b-m_s-m_d)}
{(m_s+m_d)(m_b^2-m_b m_s-m_d^2)}}}

\newcommand{\sy}{
\sqrt{
\frac{m_d(m_b-m_d)}{m_b^2-m_b m_s -m_d^2}}}

\newcommand{\sz}{
\sqrt{
\frac{m_d^2 m_s}{(m_b-m_s)(m_b^2-m_d^2)}}}

\newcommand{\ssxy}{
\sqrt{
\frac{m_u^2(m_t^2-m_u^2)(m_t-m_c-m_u)}
{(m_c+m_u)(m_t^2-m_t m_c -m_u^2)^2}}}

\begin{document}
\draft
\title{{Lepton and Quark Mass Matrices}}

\author{H. NISHIURA}
\address{
Department of General Education, 
Junior College of Osaka Institute of Technology, \\
Asahi-ku, Osaka,535-8585 Japan}
\author{K. MATSUDA and T. FUKUYAMA}
\address{
Department of Physics, 
Ritsumeikan University, Kusatsu, Shiga, 525-8577 Japan}

\date{February 17, 1999}
\maketitle

\begin{abstract}

We propose a model that all quark and lepton mass matrices have the 
same zero texture.
Namely their (1,1), (1,3) and (3,1) components are zeros.
The mass matrices are classified into two types I and II.
Type I is consistent with the experimental data in quark sector.
For lepton sector, if seesaw mechanism is not used, Type II 
allows a large \(\nu_\mu\)-\(\nu_\tau\) 
mixing angle. 
However, severe compatibility with all neutrino oscillation 
experiments forces us to use the seesaw mechanism. 
If we adopt the seesaw mechanism, 
it turns out that Type I instead of II can be consistent with 
experimental data in the lepton sector too.
\end{abstract}

One of the ultimate goals in particle physics is 
to construct the unified model of quarks and leptons.
Phenomenological construction of quark and lepton mass matrices can 
be an important step toward this goal, which reproduces and predicts 
direct and indirect observed  quantities like quark and lepton masses, 
mixing angles and \(CP\) violating phases.
In this paper we propose a model 
that all quark and lepton mass matrices, 
\(M_u\), \(M_d\), \(M_\nu\) and  \(M_e\), 
(mass matrices of up quarks (\(u,c,t\)), down quarks (\(d,s,b\)), 
neutrinos (\(\nu_e,\nu_\mu,\nu_\tau\)) and 
charged leptons (\(e,\mu,\tau\)), 
respectively) 
have the same zero texture \cite{Ramond}. 
Here \(M_\nu=-M_D^T M_R^{-1} M_D\) is 
the mass matrix of light Majorana neutrinos, 
which is considered to be constructed via the seesaw mechanism \cite{Yanagida} 
from the neutrino mass matrix,
\begin{equation}
\left(
	\begin{array}{cc}
	0 & M_D^T \\
	M_D & M_R 
	\end{array}
\right),
\end{equation}
where \(M_D\) is the Dirac neutrino mass matrix and \(M_R\) is the Majorana
mass matrix of the right-handed components. 
\(M_D\) and \(M_R\) are furthermore assumed to have the same zero 
texture matrix as \(M_\nu\).  
This assumption restricts the texture forms as follows.
\begin{eqnarray}
&&\left(
	\begin{array}{ccc}
	0 & * & 0\\
	{*} & * & *\\
	0 & * & *
	\end{array}
\right), \qquad
\left(
	\begin{array}{ccc}
	0 & 0 & *\\
	0 & * & *\\
	{*} & * & *
	\end{array}
\right), \qquad
\left(
	\begin{array}{ccc}
	{*} & 0 & *\\
	0 & 0 & *\\
	{*} & * & *
	\end{array}
\right), \qquad
\left(
	\begin{array}{ccc}
	{*} & 0 & 0\\
	0 & * & *\\
	0 & * & *
	\end{array}
\right), \qquad
\left(
	\begin{array}{ccc}
	{*} & 0 & *\\
	0 & * & 0\\
	{*} & 0 & *
	\end{array}
\right),
\nonumber\\
&&\left(
	\begin{array}{ccc}
	{*} & * & 0\\
	{*} & * & *\\
	0 & * & 0
	\end{array}
\right), \qquad
\left(
	\begin{array}{ccc}
	{*} & * & *\\
	{*} & * & 0\\
	{*} & 0 & 0
	\end{array}
\right), \qquad
\left(
	\begin{array}{ccc}
	{*} & * & *\\
	{*} & 0 & 0\\
	{*} & 0 & *
	\end{array}
\right), \qquad
\left(
	\begin{array}{ccc}
	{*} & * & 0\\
	{*} & * & 0\\
	0 & 0 & *
	\end{array}
\right).
\end{eqnarray}
Here \(*\)'s indicate suitable nonzero numbers. Among these forms we choose 
the first one because it is most close to the NNI form \cite{Branco} 
in which (2,2) component is also zero. 
Namely, our texture of mass matrix is
\begin{equation}
\left(
	\begin{array}{ccc}
	 0  & * &  0 \\
	 {*}  & * &  *\\
	 0  & * &  *
	\end{array}
\right).
\end{equation}
Indeed, this matrix leaves its form in the seesaw mechanism as
\begin{equation}
\underbrace{\left(
	\begin{array}{ccc}
	 0  & * &  0 \\
	 {*}  & * &  * \\
	 0  & * &  *
	\end{array}
\right)^T}_{\mbox{{\normalsize \(M_D^T\)}}}
\underbrace{
\left(
	\begin{array}{ccc}
	 0  & * &  0 \\
	 {*}  & * &  * \\
	 0  & * &  *
	\end{array}
\right)^{-1}}_{\mbox{{\normalsize \(M_R^{-1}\)}}}
\underbrace{
\left(
	\begin{array}{ccc}
	 0  & * &  0 \\
	 {*}  & * &  * \\
	 0  & * &  *
	\end{array}
\right)}_{\mbox{{\raisebox{-0.35cm}{\normalsize \(M_D\)}}}}
=
\left(
	\begin{array}{ccc}
	 0  & * &  0 \\
	 {*}  & * &  * \\
	 0  & * &  *
	\end{array}
\right).
\end{equation}
The nonvanishing (2,2) component distinguishes our form from NNI.
This difference, as will be shown, makes it possible to treat quark and 
lepton mass matrices universally and consistently with experiments.

Now we assign quark and lepton mass matrices as follows.
\begin{eqnarray}
&&M_{u}=
\left(
	\begin{array}{ccc}
	 0    & A_{u} & 0 \\
	A_{u} & B_{u} & C_{u}\\
	 0    & C_{u} & D_{u}
	\end{array}
\right), 
\qquad
M_{\nu}=
\left(
	\begin{array}{ccc}
	 0  & A_{\nu} &  0 \\
	A_{\nu} & B_{\nu} & C_{\nu}\\
	 0  & C_{\nu} & D_{\nu}
	\end{array}
\right),
\nonumber \\
&&M_{d}=
P_d \left(
	\begin{array}{ccc}
	 0  & A_{d} &  0 \\
	A_{d} & B_{d} & C_{d}\\
	 0  & C_{d} & D_{d}
	\end{array}
\right) P_d^\dagger 
=
\left(
	\begin{array}{ccc}
	 0  & A_{d} e^{i\alpha_{12}} &  0 \\
	A_{d} e^{-i\alpha_{12}} & B_{d} & C_{d} e^{i\alpha_{23}}\\
	 0  & C_{d} e^{-i\alpha_{23}} & D_{d}
	\end{array}
\right),\nonumber\\
&&M_{e}=
P_e \left(
	\begin{array}{ccc}
	 0  & A_{e} &  0 \\
	A_{e} & B_{e} & C_{e}\\
	 0  & C_{e} & D_{e}
	\end{array}
\right) P_e^\dagger 
=
\left(
	\begin{array}{ccc}
	 0  & A_{e} e^{i\beta_{12}} &  0 \\
	A_{e} e^{-i\beta_{12}} & B_{e} & C_{e} e^{i\beta_{23}}\\
	 0  & C_{e} e^{-i\beta_{23}} & D_{e}
	\end{array}
\right).
\end{eqnarray}
where \(P_d\equiv\mbox{diag}(e^{i\alpha_1},e^{i\alpha_2},e^{i\alpha_3})\), 
\(\alpha_{ij}\equiv \alpha_i-\alpha_j\), and 
 \(P_e\equiv\mbox{diag}(e^{i\beta_1},e^{i\beta_2},e^{i\beta_3})\), 
\(\beta_{ij}\equiv \beta_i-\beta_j\).

Let us discuss the relations between  the following texture's components 
of mass matrix \(M\),
\begin{equation}
{\normalsize M=}\left(
	\begin{array}{ccc}
	0 & A & 0\\
	A & B & C\\
	0 & C & D
	\end{array}
\right)
\end{equation}
and its eigen mass $m_i$. They satisfy
\begin{eqnarray}
B+D&=& m_1+m_2+m_3, \nonumber \\
BD-C^2-A^2&=&m_1m_2+m_2m_3+m_3m_1, \nonumber \\
DA^2&=&-m_1m_2m_3.\label{eq99020301}
\end{eqnarray}
Therefore, mass matrix is classified into two types by 
choosing \(B\) and \(D\) as follows:
\begin{eqnarray}
&&\mbox{[Type I](\(B\):large)} \quad
\ B=m_2,\ D=m_3+m_1\nonumber\\
&&\mbox{[Type II](\(B\):small)} \quad
\ B=m_1, \ D=m_3+m_2 \label{99021601}
\end{eqnarray}
Here we don't accept the case of \(B=m_1+m_2\) and \(D=m_3\) 
since in this case \(C\) becomes zero and this matrix is out of our 
texture any more. 
We adopt Type I for quark mass matrices.
For lepton sector we adopt Type I and Type II mass matrices for the case
with and without seesaw mechanism, respectively.
We proceed to discuss in detail.\\

Let us discuss the quark sector first.  
The mass matrices of Type I (\(B=m_2\), \(D=m_3+m_1\)) explains 
the quark sector consistently as will be shown. 
Assigning a definite value 
\(B=m_2\) and \(D=m_3+m_1\) in (\ref{eq99020301}) for Type I, 
we obtain
\begin{equation}
A=\sqrt{\frac{(-m_1)m_2m_3}{m_3+m_1}}, \qquad
C=\sqrt{\frac{(-m_1)m_3(m_3-m_2+m_1)}{m_3+m_1}}.
\end{equation}
Then mass matrix of Type I becomes
\begin{eqnarray}
&&M=\left(
	\begin{array}{ccc}
	0&\scriptstyle{\sqrt{\frac{m_1m_2m_3}{m_3-m_1}}}&0\\
	\scriptstyle{\sqrt{\frac{m_1m_2m_3}{m_3-m_1}}}
	& m_2 & 
    \scriptstyle{\sqrt{\frac{m_1m_3(m_3-m_2-m_1)}{m_3-m_1}}}\\
	0& \scriptstyle{\sqrt{\frac{m_1m_3(m_3-m_2-m_1)}{m_3-m_1}}}& 
    m_3-m_1
	\end{array}
\right) \simeq 
\left(
	\begin{array}{ccc}
	0& \sqrt{m_1m_2}& 0\\
	\sqrt{m_1m_2} & m_2 & \sqrt{m_1m_3}\\
	0& \sqrt{m_1m_3} & m_3-m_1
	\end{array}
\right) \nonumber \\
&&\hspace{10cm} (\mbox{for }m_3 \gg m_2 \gg m_1). \label{eq011601}
\end{eqnarray}
Here we have transformed \(m_1\) into \(-m_1\)  by rephasing.
\(M\) is diagonalized by an orthogonal matrix \(O\) as
\begin{equation}
O^T
\left(
	\begin{array}{ccc}
	0& \sqrt{m_1m_2}& 0\\
	\sqrt{m_1m_2} & m_2 & \sqrt{m_1m_3}\\
	0& \sqrt{m_1m_3} & m_3-m_1
	\end{array}
\right)
O=
\left(
	\begin{array}{ccc}
	-m_1 & 0 & 0\\
	0 & m_2 & 0\\
	0 &  0 & m_3
	\end{array}
\right),
\end{equation}
with
\begin{eqnarray}
O& = &
\left(
\begin{array}{ccc}
{\sqrt{\frac{m_2m_3^2}{(m_2+m_1)(m_3^2-m_1^2)}}}&
{\sqrt{\frac{m_1m_3(m_3-m_2-m_1)}{(m_2+m_1)(m_3-m_2)(m_3-m_1)}}}&
{\sqrt{\frac{m_1^2m_2}{(m_3-m_2)(m_3^2-m_1^2)}}} \\
-{\sqrt{\frac{m_1m_3}{(m_2+m_1)(m_3+m_1)}}}&
{\sqrt{\frac{m_2(m_3-m_2-m_1)}{(m_2+m_1)(m_3-m_2)}}}&
{\sqrt{\frac{m_1m_3}{(m_3-m_2)(m_3+m_1)}}} \\
{\sqrt{\frac{{{m_1}^2}(m_3-m_2-m_1)}{(m_2+m_1)(m_3^2-m_1^2)}}}&
-{\sqrt{\frac{m_1m_2m_3}{(m_3-m_2)(m_2+m_1)(m_3-m_1)}}}&
{\sqrt{\frac{(m_3)^2(m_3-m_2-m_1)}{(m_3^2-m_1^2)(m_3-m_2)}}} 
\end{array}
\right) \nonumber \\
&\simeq&
\left(
	\begin{array}{ccc}
 	1& \displaystyle{\sqrt{\frac{m_1}{m_2}}}&
	 \scriptstyle{\sqrt{\frac{m_1^2m_2}{m_3^3}}}\\
	\displaystyle{-\sqrt{\frac{m_1}{m_2}}}
	& 1 & \displaystyle{\sqrt{\frac{m_1}{m_3}}}\\
	\displaystyle{\sqrt{\frac{{\scriptstyle{m_1^2}}}{m_2m_3}}}
	& \displaystyle{-\sqrt{\frac{m_1}{m_3}}} & 1
	\end{array}
\right) \qquad (\mbox{for }m_3 \gg m_2 \gg m_1).
\end{eqnarray}

The mass matrices for quarks, \(M_d\) and \(M_u\) are
assumed to be of Type I as follows
\begin{equation}
M_d \simeq
P_d
\left(
	\begin{array}{ccc}
	0              &  \sqrt{m_d m_s}  &  0\\
	\sqrt{m_d m_s} &  m_s               &  \sqrt{m_d m_b}\\
	0              &  \sqrt{m_d m_b}  &  m_b-m_d
	\end{array}
\right)P_d^\dagger, \quad
M_u \simeq
\left(
	\begin{array}{ccc}
	0              &  \sqrt{m_u m_c}  &  0\\
	\sqrt{m_u m_c} &  m_c               &  \sqrt{m_u m_t}\\
	0              &  \sqrt{m_u m_t}  &  m_t-m_u
	\end{array}
\right)
\end{equation}
where \(m_d\), \(m_s\) and \(m_b\) are down quark masses and 
\(m_u\), \(m_c\) and \(m_t\) are up quark masses.
Those \(M_d\) and \(M_u\) are diagonalized by 
matrices \(P_d O_d\) and \(O_u\), respectively. 
Here orthogonal matrices \(O_d\) and \(O_u\) which diagonalize
\(P_d^\dagger M_d P_d\) and \(M_u\) are obtained from Eq. (\ref{eq990114}) 
by replacing \(m_1\), \(m_2\), \(m_3\) by \(m_d\), \(m_s\), \(m_b\) 
and by \(m_u\), \(m_c\), \(m_t\), respectively.
In this case, the Cabbibo-Kobayashi-Maskawa (CKM) \cite{CKM} 
quark mixing matrix 
\(V\) can be written as
\begin{equation}
V=P_q^{-1} P_d^{-1} O_u^T P_d O_d P_q 
\simeq
\left(
	\begin{array}{ccc}
	 |V_{11}| & |V_{12}| & |V_{13}| e^{-i \phi}\\
	-|V_{12}| & |V_{22}| & |V_{23}|\\
	 |V_{12}V_{23}|-|V_{13}|e^{i \phi}\ & -|V_{23}| & |V_{33}|\\
	\end{array}
\right). \label{eq99012501}
\end{equation}
where the \(P_d^{-1}\) factor is included to put \(V\) in the form
with diagonal elements real to a good approximation.
Furthermore, the \(P_q^{-1}\) and 
\(P_q=\mbox{diag}(e^{i\phi_1},e^{i\phi_2},e^{i\phi_3})\) with 
\(\phi_1-\phi_2=\mbox{arg}(P_d^{-1} O_u^T P_d O_d)_{12}\) and 
\(\phi_1-\phi_3=\mbox{arg}(P_d^{-1} O_u^T P_d O_d)_{23}\) are
for the choice of phase convention as Eq. (\ref{eq99012501}).
The explicit forms and numerical center values of 
components of \(V\) are 
\begin{eqnarray}
|V_{12}|&=&
\left|
\sx-\ssx e^{-i\alpha_{12}}\right. \nonumber \\
&& \left. -\ssxy \sy e^{-i \alpha_{13}}
\right| \nonumber \\
&\simeq& \left|
\sqrt{\frac{m_d}{m_s}}-\sqrt{\frac{m_u}{m_c}}e^{-i\alpha_{12}}
\right|=0.17 \sim 0.28,
\nonumber \\
|V_{23}|&=&
\left|
\ssx \sz e^{i \alpha_{12}} \right. \nonumber\\
&& \left. + \sy - \ssy e^{-i \alpha_{23}} \right| \nonumber \\
&\simeq&
\left|
\sqrt{\frac{m_d}{m_b}}-\sqrt{\frac{m_u}{m_t}}e^{-i\alpha_{23}}
\right|=0.036 \sim 0.043,
\nonumber \\
|V_{13}|&=&
\left|
\sz +\ssxy e^{-\alpha_{13}} \right. \nonumber\\
&& \left. - \ssx \sy e^{-i \alpha_{12}} \right| \nonumber \\
&\simeq & 
\left|
\sqrt{\frac{m_d^2 m_s}{m_b^3}}-
\sqrt{\frac{m_u}{m_c}}
\left(\sqrt{\frac{m_d}{m_b}}-\sqrt{\frac{m_u}{m_t}}e^{-i\alpha_{23}}
\right)e^{-i\alpha_{12}}
\right| = 0.0021\sim 0.0025, \nonumber\\
\cos\phi &\simeq &
\frac{|V_{13}|^2+|V_{12}|^2|V_{23}|^2-|V_{31}|^2}
{2|V_{12}||V_{23}||V_{13}|}
\simeq
\frac{|V_{12}|^2+m_u/m_c-m_d/m_s}{2|V_{12}|\sqrt{m_u/m_c}}=-1 \sim 1.
\label{eq122003}
\end{eqnarray}
Here we have used the running quark mass at \(\mu=m_Z\) \cite{Fusaoka}:
\begin{equation}
\begin{array}{lll}
m_u(m_Z)=2.33^{+0.42}_{-0.45}\mbox{MeV},& 
m_c(m_Z)=677^{+56}_{-61}\mbox{MeV},& 
m_t(m_Z)=181 \pm 13\mbox{GeV},\\
m_d(m_Z)=4.69^{+0.60}_{-0.66}\mbox{MeV},& 
m_s(m_Z)=93.4^{+11.8}_{-13.0}\mbox{MeV},& 
m_b(m_Z)=3.00 \pm 0.11\mbox{GeV}.
\end{array}
\label{eq123103}
\end{equation}
Let us compare (\ref{eq122003}) with the experimental values \cite{PDG}:
\begin{eqnarray}
&&|V_{12}|_{\mbox{{\tiny exp}}}=0.217 \sim 0.224, \quad 
|V_{23}|_{\mbox{{\tiny exp}}}=0.036 \sim 0.042, \nonumber\\
&&|V_{13}|_{\mbox{{\tiny exp}}}=0.0018 \sim 0.0045, \quad
 (90\%\mbox{CL}).\label{eq99012602}
\end{eqnarray}
It is remarkable that the very heavy top quark mass raises no inconsistency 
in our model. 
The reason is as follows. 
In \(|V_{23}|\), the first term of right-hand side in Eq. (\ref{eq122003})
(\(\sqrt{m_d/m_b}=0.034\)) is 
nearly equal to the experimental value 
(\(|V_{23}|_{\mbox{{\tiny exp}}}=0.036 \sim 0.042\)),
so heavy top quark mass does not make any trouble.
Whereas, in the case of Type II and also Fritzsch model \cite{Fritzsch}, 
the first term of $V_{23}$ becomes \(\sqrt{m_s/m_b}=0.18\).
So, in order to adjust to the experimental value,
the second term must be of the same order as the first term to 
cancel a large part of the first term. 
Thus top quark could not have very heavy mass.

If we adopt only the central values of quark masses in Eq. (\ref{eq123103}),
compatibility of our prediction Eq. (\ref{eq122003}) with 
the experimental values Eq. (\ref{eq99012602}) 
imposes some constraints on \(\alpha_{ij}\). 
They are depicted in FIG. 1 in the shaded strip in 
\(\alpha_{13}\)-\(\alpha_{23}\) plane.
In this figure we have superimposed the rephasing invariant Jarlskog parameter
\(J\) of quark sector, \(J=\mbox{Im}(V_{12} V_{22}^* V_{13}^* V_{23})\) 
\cite{Jarlskog}. 
However these restrictions are very sensitive to the errors of mass values 
and are not affirmative at least at this stage.
Contours represent the value of \(J\) from \(-2.3 \times 10^{-5}\) to
\(2.3\times 10^{-5}\). 
The above restriction on \(\alpha_{ij}\), therefore, gives the bound on
\(J\) as,
\begin{equation}
1.6 \times 10^{-5} \lsim |J| \lsim  2.2 \times 10^{-5}.
\end{equation}
Using the popular approximation due to Wolfenstein \cite{Wolfenstein},
the CKM quark mixing matrix can be written in terms of 
only four real parameters:
\begin{equation}
\left(
	\begin{array}{ccc}
	  {V_{11}} &  {V_{12}} &  {V_{13}} \\
      {V_{21}} &  {V_{22}} &  {V_{23}} \\ 
      {V_{31}} &  {V_{32}} &  {V_{33}}
	\end{array}
\right)
\simeq 
\left(
	\begin{array}{ccc}
     1-{\lambda^2 \over 2} & \lambda &  A\lambda^3 (\rho - i\eta) \\
     -\lambda & 1-{\lambda^2 \over 2} &  A\lambda^2 \\
     A\lambda^3 (1-\rho - i\eta) &  -A\lambda^2 &      1 
    \end{array}
\right).
\end{equation}
The measurement of the $\rho$ and $\eta$ parameters
is usually associated to the determination of the only unknown vertex 
of a triangle in the $\rho-\eta$ plane whose other 
two vertices are in (0,0) and (1,0). 
This triangle is called the unitarity triangle.
Changing freely \(\alpha_{13}\) and \(\alpha_{23}\) in Eq. (\ref{eq122003}), 
the predicted points sweep out light and dark gray regions (FIG. 2). 

Next let us discuss the lepton sector. 
We develop our arguments first without seesaw mechanism.
The mass matrix of leptons are assumed to be of Type II.
Assigning \(B=m_1\) and \(D=m_3+m_2\) (Type II) in 
Eq. (\ref{99021601}), 
we obtain from Eq. (\ref{eq99020301})
\begin{equation}
A=\sqrt{\frac{m_1(-m_2)m_3}{m_3+m_2}}, \hspace{3cm}
C=\sqrt{\frac{(-m_2)m_3(m_3+m_2-m_1)}{m_3+m_2}}.
\end{equation}
Then, we obtain the mass matrix \(M\) of Type II and 
the orthogonal matrix \(O\) which diagonalize it,
which are expressed in terms of mass eigen value \(m_i\) as 
\begin{eqnarray}
M&=&\left(
	\begin{array}{ccc}
	0&\scriptstyle{\sqrt{\frac{m_1m_2m_3}{m_3-m_2}}}&0\\
	\scriptstyle{\sqrt{\frac{m_1m_2m_3}{m_3-m_2}}}
	& m_1 & 
\scriptstyle{\sqrt{\frac{m_2m_3(m_3-m_2-m_1)}{m_3-m_2}}}\\
	0& \scriptstyle{\sqrt{\frac{m_2m_3(m_3-m_2-m_1)}{m_3-m_2}}}& 
m_3-m_2
	\end{array}
\right) \nonumber \\
&\simeq &
\left(
	\begin{array}{ccc}
	0& \sqrt{m_1m_2}& 0\\
	\sqrt{m_1m_2}
	& m_1 & \sqrt{m_2m_3}\\
	0& \sqrt{m_2m_3} & m_3-m_2
	\end{array}
\right),\\
O &=&
\left(
\begin{array}{ccc}
{\sqrt{\frac{m_2m_3(m_3-m_2-m_1)}{(m_1+m_2)(m_3-m_1)(m_3-m_2)}}}&
{\sqrt{\frac{m_1\ m_3^2}{(m_1+m_2)(m_3^2-m_2^2)}}}&
{\sqrt{\frac{m_2^2\ m_1}{(m_3-m_1)(m_3^2-m_2^2)}}} \\
{\sqrt{\frac{m_1(m_3-m_2-m_1)}{(m_1+m_2)(m_3-m_1)}}}&
-{\sqrt{\frac{m_2m_3}{(m_1+m_2)(m_3+m_2)}}}&
{\sqrt{\frac{m_2m_3}{(m_3-m_1)(m_3+m_2)}}} \\
-{\sqrt{\frac{m_2m_1\ m_3}{(m_3-m_1)(m_1+m_2)(m_3-m_2)}}}&
{\sqrt{\frac{m_2^2(m_3-m_1-m_2)}{(m_1+m_2)(m_3^2-m_2^2)}}}&
{\sqrt{\frac{(m_3)^2(m_3-m_1-m_2)}{(m_3-m_1)(m_3^2-m_2^2)}}} 
\end{array}
\right) \nonumber \\
&\simeq&
\left(
	\begin{array}{ccc}
 	1& \displaystyle{\sqrt{\frac{m_1}{m_2}}}&
	 \scriptstyle{\sqrt{\frac{m_1m_2^2}{m_3^3}}}\\
	\displaystyle{\sqrt{\frac{m_1}{m_2}}}
	& -1 & 
	\displaystyle{\sqrt{\frac{m_2}{m_3}}}\\
	\displaystyle{-\sqrt{\frac{m_1}{m_3}}}
	& \displaystyle{\sqrt{\frac{m_2}{m_3}}}
	& 1
	\end{array}
\right)\qquad(\mbox{for }m_3 \gg m_2 \gg m_1) \label{eq990114}
\end{eqnarray}
with 
\begin{equation}
O^T M O =
\left(
	\begin{array}{ccc}
	m_1 &   0  &  0  \\
	 0  & -m_2 &  0  \\
	 0  &   0  & m_3 \\
	\end{array}
\right),
\end{equation}
where we have transformed \(m_2\) into \(-m_2\).
The component (2,3) and (3,2) of \(O\) is not small comparing 
with \(\sqrt{m_1/m_3}\) in Type I. 
Therefore, due to this large mixing, 
Type II can be consistent with 
the large \(\nu_\mu\)-\(\nu_\tau\) mixing angle solution 
in atmospheric neutrino experiment as shown later.

The mass matrices of charged leptons and neutrinos are 
assumed to be of Type II as follows
\begin{equation}
M_e \simeq
P_e \left(
	\begin{array}{ccc}	
	0 & \sqrt{m_e m_\mu} & 0\\
	\sqrt{m_e m_\mu}  & m_e & \sqrt{m_\mu m_\tau}\\
	0 & \sqrt{m_\mu m_\tau} & m_\tau - m_\mu
	\end{array}
\right)P_e^\dagger,\quad
M_\nu \simeq
\left(
	\begin{array}{ccc}
	0 & \sqrt{m_1 m_2} & 0\\
	\sqrt{m_1 m_2}  & m_1 & \sqrt{m_2 m_3}\\
	0 & \sqrt{m_2 m_3} & m_3 - m_2
	\end{array}
\right),
\end{equation}
where \(m_e\), \(m_\mu\) and \(m_\tau\) are charged lepton masses and
\(m_1\), \(m_2\) and \(m_3\) are neutrino masses.
Those \(M_e\) and \(M_\nu\) are diagonalized by matrices \(P_e O_e\) and
\(O_\nu\), respectively. Here orthogonal matrix \(O_\nu\) is 
obtained from Eq. (\ref{eq990114}) with taking \(m_i\) as neutrino mass 
and \(O_e\) by replacing \(m_1\), \(m_2\), \(m_3\) 
by \(m_e\), \(m_\mu\), \(m_\tau\).
In this case, lepton mixing matrix \(U\) 
(hereafter we call it 
the Maki-Nakagawa-Sakata (MNS) mixing matrix \cite{MNS}), 
is given by
\begin{equation}
U=P_l^\dagger P_e^\dagger O_e^T P_e O_\nu P_l=
 \left(
 	\begin{array}{ccc}
 	U_{11} & U_{12} & U_{13}\\
 	U_{21} & U_{22} & U_{23}\\
 	U_{31} & U_{32} & U_{33}
 	\end{array}
 \right), \label{123002}
\end{equation}
where \(P_l=\mbox{diag}(1,i,1)\) is included to have positive neutrino mass.
\(P_l^\dagger P_e^\dagger\) factor leads \(U\) to the form whose diagonal
elements are real to a good approximation. 
We obtain the expressions of some elements of \(U\) as follows,
\begin{eqnarray}
&&U_{12}\simeq
i \left(
\sqrt{\frac{m_1}{m_2}}-\sqrt{\frac{m_e}{m_\mu}}e^{i \beta_{12}}
\right),
\qquad
U_{23}\simeq
-i \left(
-\sqrt{\frac{m_2}{m_3}}+\sqrt{\frac{m_\mu}{m_\tau}}e^{i \beta_{23}}
\right),
\nonumber\\
&&U_{13}\simeq
\sqrt{\frac{m_e}{m_\mu}}e^{i \beta{12}}
\left(
\sqrt{\frac{m_2}{m_3}}-\sqrt{\frac{m_\mu}{m_\tau}} e^{i\beta_{23}}
\right).
\end{eqnarray}
For example, substituting the neutrino masses,
\begin{equation}
m_1=1.4\times10^{-4}\mbox{eV},\quad m_2=3.2\times10^{-3}\mbox{eV},
\quad m_3=7.1\times10^{-2}\mbox{eV}, \label{eq99011501}
\end{equation}
and the charged lepton masses, \(m_e=0.51\)MeV, \(m_\mu=106\)MeV, 
\(m_\tau=1.77\)GeV, 
into Eqs. (\ref{123002}) we obtain
\begin{equation}
|U_{12}|=0.14\sim 0.28, \qquad |U_{23}|=0.033\sim 0.46, 
\qquad |U_{13}| = 0.023\sim 0.032.
\end{equation}
Here we have used
\(\Delta m_{\mbox{{\tiny atm}}}=m_3^2-m_2^2=5.0\times 10^{-3}\mbox{eV}^2\) and 
\(\Delta m_{\mbox{{\tiny solar}}}=m_2^2-m_1^2=1.0\times 10^{-5}\mbox{eV}^2\) 
with the assumption that \(m_1 \ll m_2 \ll m_3\) and \(m_1/m_2=m_2/m_3\).
Let us compare this prediction with the experimental values \cite{fukuyama}:
\begin{equation}
|U_{12}|_{\mbox{{\tiny exp}}}=0 \sim 0.71, 
\quad |U_{23}|_{\mbox{{\tiny exp}}}=0.52\sim 0.87, 
\quad |U_{13}|_{\mbox{{\tiny exp}}}=0\sim 0.22.\label{eq123101}
\end{equation}
Here we have combined the constraints 
from the recent CHOOZ reactor experiment \cite{chooz} and 
the Super KAMIOKANDE atmospheric neutrino experiment \cite{skamioka}.

Though the lepton mass matrices \(M_e\) and \(M_\nu\) of Type II lead to large
\(\nu_\mu\)-\(\nu_\tau\) mixing, 
\(|U_{23}|\) is still small compared with the experimental value.
This trouble is resolved via seesaw mechanism. 
In the seesaw mechanism, we have additional free parameters even in our model.
So we set the following assumptions guided by 
the atmospheric neutrino oscillation experiments, 
which lead to a fairly large \(\nu_\mu\)-\(\nu_\tau\) mixing.
\begin{description}
	\item[(a)] Mass matrices \(M_e\), \(M_D\) and \(M_R\) belong to Type I, 
	instead of Type II, similarly to quark mass matrices.
	\item[(b)] Mass eigen values of \(M_D\) and \(M_R\) satisfy
	\begin{eqnarray}
	m_{D3}:m_{D2}:m_{D1}=1:x:x^2, \label{eq021004}\\
	m_{R3}:m_{R2}:m_{R1}=1:x^2:x^3. 
	\end{eqnarray}
	Here \(m_{Di}\) and \(m_{Ri}\) are eigen values of 
	\(M_D\) and \(M_R\), respectively, and \(x\) is a small parameter.
\end{description}
It is noted from assumption (a) that  \(M_\nu\) itself is out of Type I via
seesaw mechanism. If we use the assumption that 
\(M_e\), \(M_D\) and \(M_R\) belong to Type II instead of Type I, 
we can not accommodate \(m_{R3}\), \(m_{R2}\) and \(m_{R1}\) to a large 
\(\nu_\mu\)-\(\nu_\tau\) mixing.
Conversely, a large mixing enforces us 
\(m_{R1}\) and \(m_{R2}\) of the same order, 
where we can not distinguish Type II from Type I.

Using assumptions (a) and (b), we obtain
\begin{eqnarray}
&&M_D \stackrel{(a)}{=}
\left(
	\begin{array}{ccc}
	0&\scriptstyle{\sqrt{\frac{m_{D1}m_{D2}m_{D3}}{m_{D3}-m_{D1}}}}&0\\
	\scriptstyle{\sqrt{\frac{m_{D1}m_{D2}m_{D3}}{m_{D3}-m_{D1}}}}
	& m_{D2} & 
    \scriptstyle{
    \sqrt{\frac{m_{D1}m_{D3}(m_{D3}-m_{D2}-m_{D1})}{m_{D3}-m_{D1}}}}\\
	0& 
	\scriptstyle{
	\sqrt{\frac{m_{D1}m_{D3}(m_{D3}-m_{D2}-m_{D1})}{m_{D3}-m_{D1}}}}& 
    m_{D3}-m_{D1}
	\end{array}
\right) \\
&& \quad \quad \stackrel{(b)}{\simeq}
m_{D3} 
\left(
	\begin{array}{ccc}
	0          & x \sqrt{x} & 0\\
	x \sqrt{x} & x          & x\\
	0          & x          & 1
	\end{array}
\right) \label{eq021001}
\end{eqnarray}
and similarly,
\begin{equation}
M_R \simeq
m_{R3}
\left(
	\begin{array}{ccc}
	0            & x^2 \sqrt{x} & 0\\
	x^2 \sqrt{x} & x^2          & x \sqrt{x} \\
	0            & x \sqrt{x}   & 1
	\end{array}
\right).
\end{equation}
Then the neutrino mass matrix \(M_\nu\) is given by
\begin{equation}
M_\nu=-M_D^T M_R^{-1} M_D=
-\frac{(m_{D3})^2}{m_{R3}}
\left(
	\begin{array}{ccc}
	0        & \sqrt{x}         & 0             \\
	\sqrt{x} & 1+(\sqrt{x}-x)^2 & 1-(\sqrt{x}-x)\\
	0        & 1-(\sqrt{x}-x)   & 1
	\end{array} 
\right).\label{eq021003}
\end{equation}
The orthogonal matrix 
which diagonalizes Eq. (\ref{eq021003}) is 
\begin{equation}
O_\nu \simeq 
\left(
	\begin{array}{ccc}
	-\frac{1}{12}
	\frac{-72+48 \sqrt{3}-9 \sqrt{x}+5 \sqrt{3} \sqrt{x}}
	{(\sqrt{3}-1)(3-\sqrt{3})^{3/2}} &
	\frac{1}{12}
	\frac{72+48 \sqrt{3}+5 \sqrt{3} \sqrt{x}+9 \sqrt{x}}
	{(1+\sqrt{3})(3+\sqrt{3})^{3/2}} &
	\frac{\sqrt{2}}{4} x^{1/2}+\frac{\sqrt{2}}{8} x \\
	\frac{1}{24}
	\frac{-72+48 \sqrt{3}+21 \sqrt{x}-7 \sqrt{3} \sqrt{x}}
	{(3-\sqrt{3})^{3/2}} &
	\frac{1}{24}
	\frac{72+48 \sqrt{3}-21 \sqrt{x}-7 \sqrt{3} \sqrt{x}}
	{(3+\sqrt{3})^{3/2}} &
	\frac{\sqrt{2}}{2}+\frac{5 \sqrt{2}}{32} x \nonumber\\
	-\frac{1}{24}
	\frac{-72+48 \sqrt{3}-15 \sqrt{x}+5 \sqrt{3} \sqrt{x}}
	{(3-\sqrt{3})^{3/2}} &
	-\frac{1}{24}
	\frac{72+48 \sqrt{3}+15 \sqrt{x}+5 \sqrt{3} \sqrt{x}}
	{(3+\sqrt{3})^{3/2}} &
	\frac{\sqrt{2}}{2} -\frac{7 \sqrt{2}}{32} x
	\end{array}
\right),
\end{equation}
And the eigen mass is
\begin{eqnarray}
&&m_1 \simeq
\frac{m_{D3}^2}{m_{R3}}
\left\{
\left(\frac{1}{2}-\frac{\sqrt{3}}{2}\right) \sqrt{x}
-\left(\frac{3}{8}-\frac{\sqrt{3}}{24}\right) x
\right\}, \nonumber\\
&&
m_2 \simeq
\frac{m_{D3}^2}{m_{R3}}
\left\{
\left(\frac{1}{2}+\frac{\sqrt{3}}{2}\right) \sqrt{x}
-\left(\frac{3}{8}+\frac{\sqrt{3}}{24}\right) x
\right\}, \quad
m_3 \simeq 
\frac{m_{D3}^2}{m_{R3}}
\left\{
2-\sqrt{x}+\frac{7}{4} x
\right\}. \label{eq021007}
\end{eqnarray}
For numerical estimation we assume that 
mass pattern Eq. (\ref{eq021004}) is same as that of up quark.
\begin{equation}
m_t(m_Z): m_c(m_Z): m_u(m_Z) =1: x: x^2, \qquad (x \simeq 0.0036)
\end{equation}
and, therefore, \(m_{D3}=k\times m_t(m_Z)\), 
\(m_{D2}=kx \times m_t(m_Z)\) 
and \(m_{D1}=kx^2 \times m_t(m_Z)\). 
Using the assumption (a) that \(M_e\) belong to type I, 
the mass ratios of light Majorana neutrinos, 
the MNS matrix \(U\) and the rephasing invariant Jarlskog parameter \(J\)
of lepton sector become
\begin{eqnarray}
&& m_3:m_2:-m_1 \simeq 1.0:0.04:0.01,\\
&&U=P_e^\dagger O_e^T P_e O_\nu
\nonumber \\
&& \quad \simeq
\left(
	\begin{array}{ccc}
   -0.88 - 0.02 e^{-i\beta_{12}} & 0.46-0.04 e^{-i\beta_{12}} &
	0.022-0.049e^{-i\beta_{12}}\\
	0.34-0.06 e^{i\beta_{12}}    &
	0.62+0.03e^{i\beta_{12}}+0.01e^{-i\beta_{23}} &
	0.71 - 0.01 e^{-i\beta_{23}}\\
   -0.31 + 0.01 e^{i\beta_{23}} & - 0.64 + 0.01e^{i \beta_{23}}  &
	0.71 + 0.01 e^{i\beta_{23}}
	\end{array}
\right) \label{eq021008}
\end{eqnarray}
and \(|J| \lsim 0.01\). 
Here we have assumed that
the changes of lepton masses and the MNS mixing 
from \(\mu=m_Z\) to \(\mu=\)MeV are very small.
At this stage only one parameter, \(m_{R3}\), 
still remains free.
It will be determined from \(\Delta m_{32}^2=5.0\times 10^{-3} eV^2\) as 
\begin{equation}
m_{R3}=k^2 \times (9.0 \times 10^{23}) \mbox{eV}.
\label{eq021005}
\end{equation}
Thus we have fixed parameters so as 
to adjust the atmospheric neutrino oscillation experiments.
The assumptions (a) and (b) are not unique and their justification is
checked by the compatibility with the solar neutrino deficit experiments.
From Eqs. (\ref{eq021007}), (\ref{eq021008}) and (\ref{eq021005}), 
we have the restrictive prediction.
\begin{equation}
\Delta m_{21}^2 \simeq 7.8 \times 10^{-6} \mbox{eV}^2, \quad
\tan^2 \varphi \equiv \frac{|U_{13}|^2}{|U_{23}|^2+|U_{33}|^2} \simeq 0, \quad
\tan^2 \omega \equiv \frac{|U_{12}|^2}{|U_{11}|^2} \simeq 0.27,
\end{equation}
which are superimposed on the analyses by Fogli et. al. \cite{Fogli} (FIG. 3).
The star indicates our prediction. 
The position of star has been determined from the atmospheric neutrino
experiments and was free from the solar neutrino deficit experiments.
Nevertheless its position in the allowed region of solar neutrino experiments.

Conclusive remarks are in order.
We started with the same type of 4 texture zero mass matrices 
both for quarks and leptons. They were classified into Type I and II.
Type I explains quark sector consistently. 
For the lepton sector Type II, on the other hand, 
reproduces qualitatively large lepton mixing. 
However, best fitting with experimental data requires the seesaw mechanism 
in lepton sector with Type I mass matrices similarly to quarks.

\ \\
Acknowledgements

We are grateful to H.Minakata and O.Yasuda for valuable discussions.
Special thanks are due to Y. Koide for his enlightening suggestions on 
phenomenological mass matrix models.

\begin{figure}[htb]
 	\begin{center}
 	\leavevmode
 	\epsfile{file=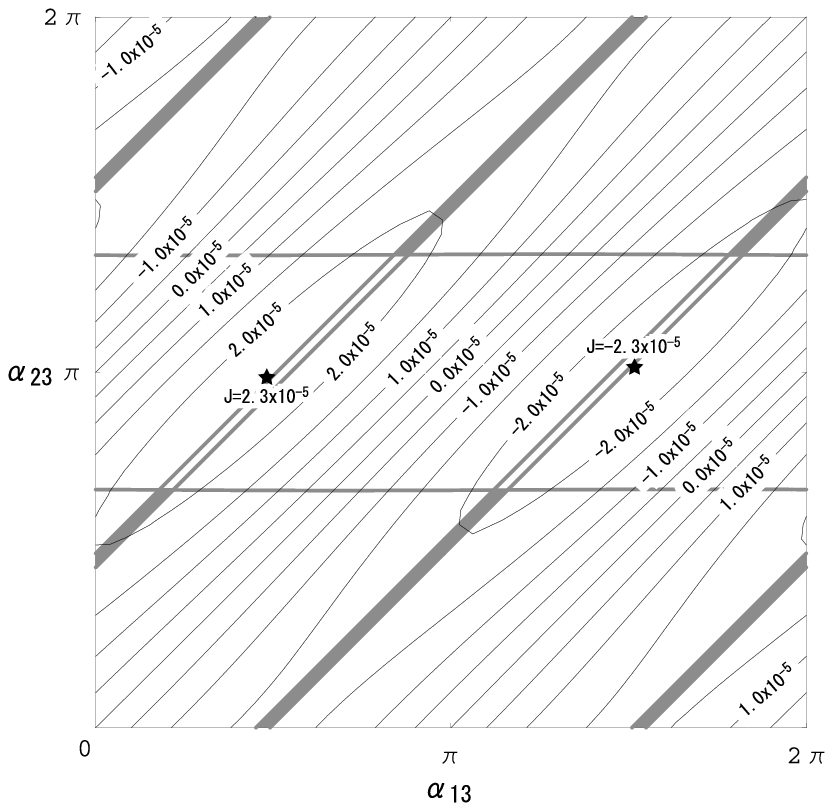}
 	\end{center}
\caption{
The allowed region on 
\(\alpha_{13}\) - \(\alpha_{23}\) plane is depicted 
by the shaded areas. In the allowed region, the contours indicate
the rephasing invariant of Jarlskog parameter 
\(J ( \equiv \mbox{Im}(V_{12} V_{22}^* V_{13}^* V_{23}))\) of quark sector.}
\label{fig1}
\end{figure}

\begin{figure}[htb]
 	\begin{center}
 	\leavevmode
 	\epsfile{file=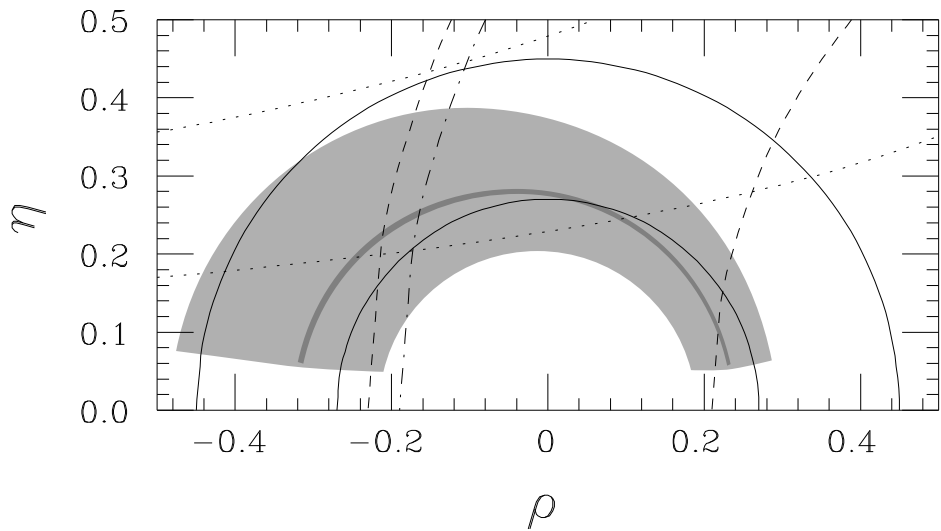}
 	\end{center}
\caption{
The vertex position of unitarity triangle predicted by our model 
is superimposed on the diagram restricted by hadron experiments.
Our predictions is obtained by changing \(\alpha_{13}\) and \(\alpha_{23}\) 
freely in Eq. (15) with no approximation.
If each quark mass takes the center values in Eqs. (16),
the dark gray region is allowed.
On the other hand, taking the error of each quark mass into consideration,
we obtain the light gray region.}
\label{fig2}
\end{figure}

\begin{figure}[htb]
 	\begin{center}
 	\leavevmode
 	\epsfile{file=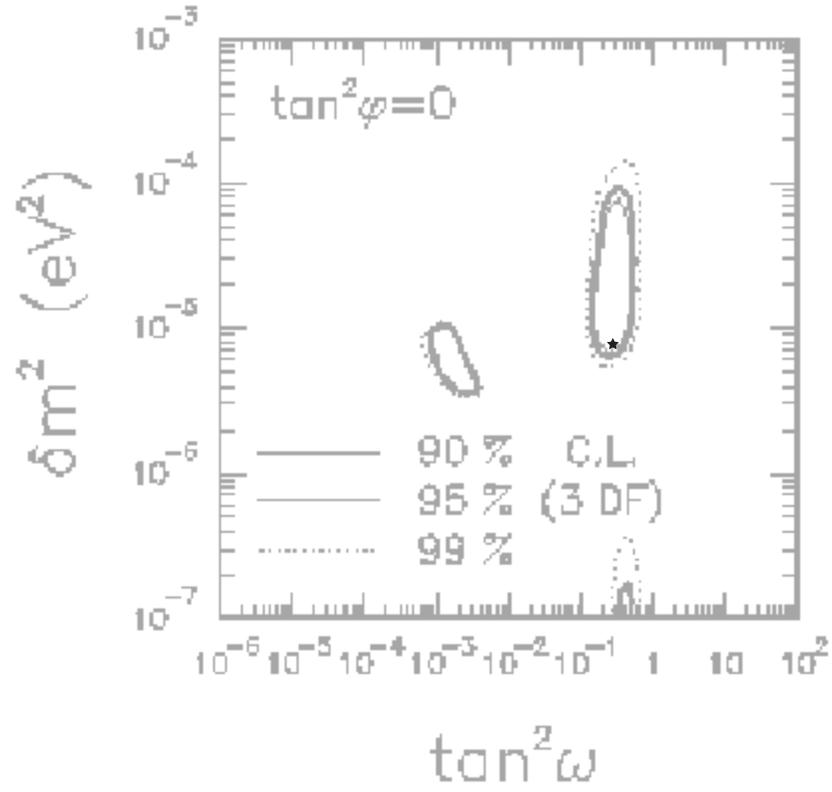}
 	\end{center}
\caption{
The solid line and dotted line show 90\% C.L. and 99\% C.L., respectively,
which were derived from the three-flavor analysis of 
the solar neutrino deficit experiments [15].
The star indicates our prediction.}
\label{fig3}
\end{figure}

\end{document}